# Broadband Searches for Extraterrestrial Technological Intelligence: A New Strategy to Find Nearby Alien Civilizations

B. Zuckerman
Department of Physics and Astronomy, University of California, Los Angeles, CA 90095-1562, USA; ben@astro.ucla.edu

## Abstract

One of the most interesting questions that astronomy can hope to answer is: Are we alone in our Milky Way Galaxy? A detection of an electromagnetic (EM) signal generated by an extraterrestrial technological intelligence, or the presence in our solar system of an alien probe, would answer this question in the negative. Purposeful interstellar communication is a two-way street—the transmitting and receiving technological intelligences (TIs) both need to do their parts. As the receiving TI, our EM search programs should incorporate a model of what a transmitting TI is likely to be doing. Published works on the search for extraterrestrial technological intelligence (SETI) have generally not done so, and thus have often been suboptimally designed. We propose an improved search technique that more closely corresponds to astronomical surveys that have been undertaken for reasons that have nothing to do with SETI. Published non-SETI radio and optical surveys are sufficiently extensive that they already supply meaningful constraints on the prevalence of nearby, purposely communicative alien civilizations. Purposeful communication can also include the sending of spaceships (probes). The absence of evidence for alien probes in the solar system suggests that no alien civilization has passed within ∼100 lt-yr of Earth during the past few billion years.

## 1. Introduction

Modern astronomy is poised to illuminate questions such as whether life exists or has ever existed elsewhere in our solar system, and whether there are worlds that orbit stars other than the Sun on which life now exists. A subset of such possibilities involves technological life in the Milky Way—has life elsewhere, capable of technology, evolved during eons of time? More than 60 yr ago proposals to communicate with alien technological life—via radio waves or at optical or infrared wavelengths (G. Cocconi & P. Morrison 1959; R. N. Schwartz & C. H. Townes 1961)—were published. Negative results of many search programs, at radio and optical wavelengths, have appeared (e.g., J. Tarter 2001; D. C. Price et al. 2020; J. L. Margot et al. 2023; G. W. Marcy & N. K. Tellis 2024; N. Tusay et al. 2024).

Published works on the radio search for extraterrestrial technological intelligence (SETI) include the implicit or explicit assumption that interstellar transmissions will usually be power-limited ("power starved"). Thus, radio search programs have employed very narrow (few Hz) bandwidths (BWs)—because, if an extraterrestrial technological intelligence (ETI) has a given (limited) amount of power to transmit, then the way to maximize the signal-to-noise ratio at the receiving antenna is to use very narrow transmission and reception BWs. However, because we, as the receiving technological intelligence (TI), do not know a priori the wavelength of a transmitted signal, detection requires complex hardware and software (D.C. Price et al. 2020; J. L. Margot et al. 2023; N. Tusay et al. 2024) to cover, simultaneously, a huge number of very narrowband channels. Even then, the total BW covered in published searches has comprised only a modest percentage of the full radio/microwave window. The assumption of power starvation has motivated SETI projects to use some of the largest radio telescopes in the world (D. C. Price et al. 2020; J. L. Margot et al. 2023).

In an early, wide-ranging paper, "Transmission of Information by Extraterrestrial Civilizations," N. S. Kardashev (1964) appraises isotropic vs high-directivity antennas: "we shall assume the [transmitted] radiation isotropic. There appears to be no point, in our view, in discussing the possibility of establishing outer space communications when high-directivity antennas are being used for transmission, since the probability of success in establishing such communications is virtually nil. The high directivity of the radiation can be conveniently used, in all probability, only after a two-way communication exchange has been set up." If transmission is isotropic, as Kardashev proposes, then a signal in any given direction will be weak, thus leading to power starvation. Our primary conclusion, based on the model outlined in Section 2, is the opposite of the conclusion reached in the preceding quote. That is, a sufficiently nearby, communicative TI will use *high-directivity antennas*, thus vitiating any concerns with power starvation. Then the most uncertain factor in our communication with a nearby ETI will not be power starvation, but rather the wavelength of transmission; this may be radio, infrared, or optical. Thus, search programs should aim to cover as much of the electromagnetic (EM) spectrum as possible—this is very difficult to do with currently designed radio SETI programs.

For nearby, *purposely* communicative, technological civilizations, power starvation—the primary motivation for use of filters having very narrow BWs—is irrelevant. Our principal assumption is that a purposely communicative technological civilization will do its *technological best* to establish communication with other ETIs. This opens the possibility of serendipitous detection of an alien transmitter in EM sky surveys undertaken for reasons that have nothing to do with SETI. If a nearby ETI seriously wants to communicate with other nearby TIs then, as shown below, it can and will transmit signals that can be detected even by a novice technological civilization such as our own that is engaged in sensitive, but run-of-the-mill, astronomical research that utilizes modestsized telescopes.

Relevant astronomical research covers radio, infrared, and optical wavelengths. During the past 100 yr, optical astronomers have had a myriad of opportunities to detect a signal from an ETI, but nothing has been reported. Evaluation of how constraining these optical nondetections are—in the context of the transmission model proposed in the present paper—is complicated (see Section 6). A proper analysis, in a SETI context, of the ensemble of 100 yr of optical spectroscopy warrants a full-length paper. That paper might well conclude that all of the hundreds of thousands of nearby solar-type stars considered below have been observed optically with sufficient sensitivity to reveal any purposely communicative technological civilization. Notwithstanding the possibility of such a strong outcome, the radio studies we focus on here are still of great importance—because an alien's choice of preferred transmission wavelength is essentially unknown.

The primary goal of the present paper is to illustrate how best to design a radio search strategy that will successfully detect any *purposely* communicative, technological civilizations currently in the solar neighborhood. Failing a detection, a quantitative limit can be placed on the number of ETIs out to some distance from the Sun. The limits to be derived apply only to ETIs that are doing their *technological best* to establish communication with other technological species in their vicinity.

Thus, the derived limits are not relevant for constraining the number of ETIs that have no interest in interstellar communication, or are "neutral" about the possible existence of other technological civilizations and are not trying, or at least not trying hard, to make their presence known. Nor do they apply to any ETI that fears that other TIs may display aggressive, negative behaviors; presumably such ETIs will try to hide their existence and will certainly not be broadcasting signals toward Earth. Likewise, they do not apply to any ETI that uses a means of communication that is not EM waves (e.g., neutrinos, gravitational waves, axions). While one cannot be certain that EM waves will be the medium of choice, all alternatives that have been mentioned in the literature are immensely more difficult to generate and/or to capture.

Arguably, for we humans, the most interesting ETI would be one that is near Earth, either now or at some point in the past. Thus, purposeful communication can also include the sending of macroscopic bodies such as spaceships (probes) to our solar system during past eons of time. We quantitatively consider the implications rendered by lack of evidence for any alien probes, most plausibly in orbit around Earth, or the aliens themselves in the solar system.

The paper is organized as follows. In Section 2 we describe a plausible (likely?) EM transmission model for a nearby, purposely communicative ETI. The received strength at Earth of an anticipated radio wavelength signal sent by such an ETI is calculated in Section 3. Sections 4 and 5 compare the Section 3 signal strengths with the sensitivities of SETI and non-SETI radio sky surveys, respectively. Section 6 concerns infrared and optical SETI. Section 7 considers travel by interstellar spaceships and describes how constraints can be placed on the number of communicative ETIs that have passed by the solar vicinity during the past few billion years. Section 8 utilizes previous sections to estimate upper limits to the number of communicative ETIs in our Milky Way
Galaxy

## 2. Electromagnetic Interstellar Transmission

Both the transmitting and receiving TIs need to consider a five-dimensional phase space that includes distance, direction, sensitivity, wavelength, and time (epoch). For distance, in a benchmark model calculation, we will assume a sphere of radius 200 pc (∼650 lt-yr) about both the receiving and transmitting societies. While there is nothing sacred about 200 pc, it illustrates the efforts that the two societies will need to make to optimize the possibility of success.

For direction we make the standard assumption that other life forms will be similar to the only life we know—us. Liquid water is essential for Earth life. Thus to find life as we know it, we must find terrestrial-size planets in the "habitable zone"—a region where water will be in a liquid state on at least part of the planet at least some of the time. But we are considering technological life that took about 4.5 billion years to develop here. So for "direction," we and the ETI search for other TIs only at old stars not too different from our Sun. This means that only planets that orbit stars of mass less than 1.25 solar masses (F6 spectral type) live long enough (4.5 billion years) for a species to develop technology (M.J. Pecaut & E. E. Mamajek 2013). We assume that the lower mass limit of interest corresponds to an M2-type star of mass 0.44 solar masses. Lower-mass M-type stars (M3–M9) are probably poor sites for long-lived habitable planets (A. Segura et al. 2010; S. Rugheimer et al. 2015; X. Li et al. 2025). Recent JWST observations of TRAPPIST-1b and c are consistent with airless, atmosphere-stripped worlds (M. Gillon et al. 2026). Even if life does originate, the build-up of atmospheric oxygen, necessary for complex life, may be exceedingly slow (J. J. Soliz & W. F. Welsh 2026).

According to a recent all-sky survey (J. D. Kirkpatrick et al. 2024), in a volume of radius 200 pc there are about 500,000 single, solar-type stars. In the calculations that follow, we will assume that 200,000 of these are sufficiently old (>4.5 billion years) to be of interest for hosting a technological civilization. Age is one of the hardest properties to measure for isolated stars, i.e., those not in a star cluster. Techniques include stellar evolutionary modeling, asteroseismology, chromospheric activity, and rotation rate (e.g., C. Pezzotti et al. 2026). Given the youth of our modern astronomy, placement of a given star into an old or not-so-old category is prone to errors. Thus, a targeted SETI search program may have to point its telescopes toward ∼300,000 nearby stars to be sure that all old ones are included. It is generally agreed that, for technological life to not be exceedingly rare, the age of a typical TI must be many millions of years, probably at least hundreds of millions of years. Thus, we will assume that any venerable transmitting ETI will have determined which solar-type stars within 200 pc are old.

Statistics from studies based on Kepler satellite data suggest that about 30% of 200,000 planetary systems will include a potentially habitable rocky planet (J. K. Zink & B. M. S. Hansen 2019; S. Bryson et al. 2021), i.e., about 60,000 potentially habitable planets. *How many of these 60,000 are actual living worlds?*

In preparation for its program of discovery and communication with another TI, the transmitting ETI will surely have investigated—with large space telescopes—all of the 200,000 relevant old star systems. Thus, they will have determined which ∼60,000 stars within 200 pc are orbited by planets in the habitable zone. But, more important, these investigations will reveal which of these 60,000 show evidence that life actually exists on a rocky planet of appropriate size.

Arguably, the most interesting—and currently unknown—quantity in the quest for life in the Universe is what fraction ("*F*") of potentially habitable planets actually develop life that then exists for a cosmically meaningful length of time—a "living world." Determination of *F* was the goal of the US Terrestrial Planet Finder and European Darwin programs that were considered, but then abandoned, decades ago. These have been reborn in the form of ESA's new mission called Large Interferometer for Exoplanets (LIFE; S. P. Quanz et al. 2022) and NASA's Habitable Worlds Observatory (V. Van Eylen et al. 2025). While we do not yet know the value of *F*, a million-year-old technological society surely does. It is inconceivable that such a mature ETI will not have constructed large space telescopes and interferometers—far more powerful than LIFE—to be used for numerous purposes, including determination of *F*.

Since we do not now know *F* we will assume that it equals 10%. This guesstimate for *F* could be too small by only a factor of a few, or too large by orders of magnitude. One factor to be considered in the determination of *F* is the probability of the origin of life. We do not understand how life originated here, but the fact that all Earth life is chiral—of one handedness—suggests that life might have originated only once in the entire long history of Earth. Many models to explain life's chirality have appeared in the literature, but the idea that life originated on Earth only once surely is the most straightforward. Thus, the origin of life may be exceedingly difficult. Then there are changing environmental conditions over billions of years. In our own solar system, of Earth, Mars, and Venus, even if life originated on all three, only Earth has evolved into a living world. Factors necessary for the longterm survival and evolution of life might be the presence of a strong planetary magnetic field and/or the presence of a large moon (J. Laskar et al. 1993), both true for Earth but not for Mars and Venus.

If $F = 10\%$, then one might think that the number of target stars that an ETI might need to transmit toward is ∼6000—the number of living worlds that orbit single main-sequence stars of appropriate mass (spectral types) within a radius of 200 pc. But 6000 is likely far greater than the number of living worlds an ETI will actually transmit toward. One reason is the requirement that a long-lived technological civilization can develop only on a planet whose surface contains extensive areas of both land and water—i.e., a true Earth analog. A planet whose surface is totally covered with water might develop life, but it is not going to develop a technological civilization; nor is a dry planet like Mars or Venus.

What percentage of 6000 living worlds has the desired mix of land and ocean areas? This fraction ("*f*") is currently unknown to us—we do not know how Earth obtained its water (e.g., C. M. Guimond et al. 2026). A long-favored model involves delivery of wet asteroids and comets during late times in Earth's formation years (Y. Hasegawa et al. 2025). A recent, very different, model—that proposes chemical reactions between a hydrogen-rich primordial atmosphere and a magma ocean as Earth formed—has garnered substantial support (E. D. Young et al. 2023; J. G. Rogers et al. 2024; F. Miozzi et al. 2025). Our interest is the fraction "*f*" of living worlds that have comparable amounts of surface land and water. Clearly, *f* remains unconstrained.

Specular reflection of host starlight from surface bodies of water—referred to as ocean glint—has been considered in numerous papers (e.g., J. Lustig-Yaeger et al. 2018; D. J. Ryan & T. D. Robinson 2022). Phase and spectral variability on rotating planets provide one method to determine *f*. Further in the future, the space antennas that discover living worlds should have sufficient spatial resolution, say 3000 km, to separate continents and oceans. For a transmitter 200 pc from Earth, an interferometer operating at a wavelength of, say, 8000 Å would require a baseline length of 2000 km. Interferometers could be positioned in space, or on large asteroids, or on airless moons (e.g., G. T. van Belle et al. 2025).

Should $f$ be around 10%, then, after surveying with their space antennas the 500,000 single, solar-type star systems within 200 pc, an ETI would end up having to transmit toward only 600 planets, one of which would be Earth. In other words, the product of $F$, $f$, and other factors listed above—all known to a venerable transmitting society thanks to their large space antennas—will make feasible a long-term transmission project that targets only a fairly small number of stars. In contrast, because we have not yet built and used such antennas, we must survey hundreds of thousands of nearby stars in any complete SETI program.

For a long-lived transmitting civilization that is moving relative to other nearby star systems, the 600 Earth-like planets with characteristics potentially supportive of technological life will change every few million years or so. So the 600 transmitting antennas will be repointed and maybe reprogrammed every once in a rare while. But transmission need never stop because 60 MW can be supplied continuously at each antenna by a square stellar power collector with side 200 m, conversion efficiency of 30% from starlight to beamed microwaves, infrared, or optical light, and orbiting at, say, 0.5 au from its solar-like central star. (Mercury orbits at ∼0.4 au from the Sun.)

For a 60 MW transmitter, broadcasting into a few Hz BW, located 200 pc from Earth that imprints a "footprint" of radius 1.4 au at our solar system, the received flux density in this narrow bandpass at Earth will be $10^{10}$ Jy (1 Jy = $10^{-26}$ W m$^{-2}$ Hz$^{-1}$). This beam radius at our solar system will ensure that Earth will always be receiving a signal, no matter where it is in its orbit around the Sun.

The transmitting ETI will need to decide how large a footprint its beam will cover at the receiving end. We assume that at our solar system the diameter of the beam sent by the ETI will be 2.8 au (only Earth is of interest to the transmitting ETI). How large a phased array transmitter will be required? Assume that transmission is at 10 GHz (0.03 m) and from a distance of 200 pc. The array diameter would be ∼1000 km. One way to reduce the array size is to allow a larger footprint at our solar system; this results in reduced signal strength at Earth. This could be (partially) compensated for by using a transmission power greater than the 60 MW assumed in our benchmark calculation. Desire for high spatial resolution highlights an advantage of transmission in the infrared (Section 6)—the diameter of the transmitting array can be much smaller.

## 3. Search Sensitivities

Major radio SETI programs, e.g., Breakthrough Listen (BL; H. Isaacson et al. 2017; B. S. Wlodarczyk-Sroka et al. 2020) and J. L. Margot et al. (2023), have typically searched between 1 and 10 GHz—the "sweet spot" where the combination of Galactic background, Earth atmospheric, and quantum noise is at a minimum—with filters a few Hz wide (J. Tarter 2001; D.C. Price et al. 2020; J. L. Margot et al. 2023). Radio SETI programs assume that an alien transmitter will be radiating isotropically—rather than only in a few hundred directions—in part because they have not adequately considered or modeled what the transmitting ETI is likely to be doing. These programs define an equivalent isotropic radiated power (EIRP) that produces a flux $S$ (W m$^{-2}$) from a transmitter a distance $R$ away:

$$S = EIRP/4\pi R^2.$$

Quoted detection sensitivities ($S$) of D.C. Price et al. (2020) and J. L. Margot et al. (2023) are about

$$10 \times 10^{-26}\ (\mathrm{W\ m^{-2}}).$$

This sensitivity in a few Hz BW is then roughly $10^{-26}$ (W m$^{-2}$ Hz$^{-1}$), i.e., 1 Jy. This sensitivity can be realized only if the wavelength of the transmitted signal, assumed to be a few Hz wide, matches the wavelength to which the receiver is tuned. Since the transmission wavelength is not known, one must tune the wavelength of reception. So, for example, if the receiver BW is 3 Hz and one wants to search over 3 GHz, the receiving wavelength must be adjusted $10^9$ times or one must utilize complex hardware and software to cover a huge number of narrow channels simultaneously (D. C. Price et al. 2020; J. L. Margot et al. 2023; N. Tusay et al. 2024).

Quoted published sensitivities of 1 Jy can be compared to the signal that would be received at Earth from our benchmark 60 MW transmitter. Such transmission is directed toward only tiny footprints on the sky rather than isotropically. As noted in Section 2, the result is a received flux density ∼$10^{10}$ Jy in a BW that is a few Hz wide. Thus, the receiver BW can be orders of magnitude larger and the alien signal will still be easily detected above the noise in the surveys described in Sections 4 and 5.

A virtue of using a few Hz BW, matched to a similarly narrow transmission BW, is that a detection immediately indicates an alien technology—because nature cannot produce such a narrow emission line.

If a broad BW is used in a search, then one forfeits this immediate advantage. However, old, solar-like stars are not detected as sources in any radio survey (L. D. Matthews 2025). Thus, any such star with detectable emission would immediately be flagged as very interesting—there is no way that a signal from an ETI in orbit at an old solar-type star would be overlooked in a broadband radio sky survey that includes the 200,000 nearby stars considered above.

## 4. Radio Wavelength SETI Searches

The first extensive radio search was carried out by Palmer and Zuckerman in the early 1970s; they observed ∼670 stars within 25 pc of Earth. This was the most extensive targeted radio SETI search during the 20th century (R. H. Gray 2021). Although a detailed list of the observed stars is no longer available, it is unlikely that all of the stars satisfied the constraints given for the transmission model described in Section 2. The channel widths of 52 and 4 kHz were much wider than the few Hz resolution favored in modern SETI searches. However, given the likely strength of a transmitted signal as seen at Earth (as noted above), these spectral resolutions were unlikely to have impaired detectability. A full list of 20th century radio and optical SETI programs may be found in J. Tarter (2001).

Recent radio wavelength searches include J. L. Margot et al. (2023), who observed 11,680 stars over the wavelength interval 1.15–1.73 GHz with the 100 m diameter Green Bank Telescope. These target stars differ substantially from the set of old nearby solar-like stars considered in Section 2. Because the mean distance to the Margot sample is ∼8000 lt-yr, it is appropriate to consider EIRP—since at such distances there is no reason to expect that an ETI will transmit preferentially toward Earth.

Similarly, the target selection for BL was largely not focused on old, nearby, solar-type stars (H. Isaacson et al. 2017). The 288,315 target stars reported in a major BL survey paper (B. S. Wlodarczyk-Sroka et al. 2020) overlaps only minimally with the model 200 pc search of 200,000 old solartype stars described in Section 2. Figure 13 in L. Manunza et al. (2025) summarizes the number of stars surveyed in a dozen SETI programs in the range 1–10 GHz during the past decade or so. Most of the Figure 13 programs covered far fewer than 200,000 stars, while the targeted stars in the few published programs that covered 1,000,000 or more stars show minimal overlap with the 200,000 stars in the model outlined in Section 2.

## 5. Non-SETI Radio Wavelength Surveys

Apparently overlooked in previous discussions of radio SETI is the relevance of extensive, sensitive radio sky surveys undertaken for reasons that have nothing at all to do with SETI. An advantage of such surveys is they are *broadband*— so much less time is consumed stepping along in wavelength. In total, non-SETI radio surveys have probably covered more of the position/wavelength space relevant to the 200 pc volume considered here than have the totality of explicit radio SETI search programs.

Taken together, high-sensitivity, centimeter-wavelength surveys have covered almost the entire sky: VLASS (M. Lacy et al. 2020), NVSS (J. J. Condon et al. 1998), FIRST (R. H. Becker et al. 1995), SUMSS (T. Murphy et al. 2007), and the Sardinia Radio Telescope (SRT) (L. Manunza et al. 2025). The sensitivities of these surveys are easily sufficient to detect a signal from the benchmark model outlined in Section 2. No detections of nearby, old, solar-type stars have been reported in any of these papers (L. D. Matthews 2025). As regards sensitivity, consider the VLASS Very Large Array 3 GHz survey (M. Lacy et al. 2020). The filter width is 2 MHz and the quoted sensitivity is better than 1 mJy. The signal anticipated in a few Hz BW in the model discussed in Section 2 is $10^{10}$ Jy, or about $10^4$ Jy if smeared over 2 MHz. For comparison, the Crab Nebula has a flux density ∼1000 Jy at radio frequencies from 1 to 10 GHz. The Crab is one of the brightest and most studied radio sources in the sky and is often used as a calibration source.

Notwithstanding their excellent sensitivity and sky coverage, the abovementioned surveys, along with a few others not listed, cover only a limited portion of the radio spectrum— VLASS from 2 to 4 GHz and the others in the vicinity of 1 GHz. Most of the radio spectral window remains to be surveyed with broadband receivers. Therefore, it would be premature to use these non-SETI surveys to place a quantitative limit on the number of nearby ETIs whose highpower directional transmissions occur at radio wavelengths.

## 6. Infrared and Optical SETI

Optical/infrared SETI is beyond the scope of the present paper and deserves a full-length paper. As noted in the published literature, for omnidirectional SETI transmissions, radio wavelengths are the best choice. But for beamed transmissions the optimal wave band is not obvious, at least not at the current state of technology (C. H. Townes 1983; B. Zuckerman 1985; A. Betz 1986).

Optical spectral surveys go back 100 yr to the Henry Draper (HD) catalog of 225,000 stars. Because HD is a brightnesslimited catalog, it contains many stars sufficiently massive to not live long enough on the main sequence to develop technological life. Still, some solar-like stars of interest here were observed more than 100 yr ago with sufficient sensitivity to detect beamed optical transmissions of the sort described above for radio transmission. In contrast, the first substantial radio survey took place only in the 1970s, and little more was done until the present century (R. H. Gray 2021).

It would be worthwhile to research published papers, both SETI focused (A. W. Howard et al. 2004; G. W. Marcy et al. 2022; G. W. Marcy & N. K. Tellis 2023, 2024; B. Fields and J. C. Goodman 2025) and general optical and infrared surveys, to determine how many of the hundreds of thousands of stars considered in Section 2 have been observed at infrared or optical wavelengths and with what spectral resolution. The Sloan Digital Sky Survey (SDSS) of most of the sky covered 3800–9200 Å with spectral resolution 1500–2500 (D. G. York et al. 2000; Abdurro'uf et al. 2022). With this spectral resolution, detection of a laser transmitter against the stellar background flux will depend on the laser power and the luminosity of the host "Sun-like" star (the fainter the better for laser detection). An F6 star is 100 times more luminous than an M2 star. The SDSS is now being extended to cover the entire sky (J. Kollmeier et al. 2026).

The authors of a paper that explores an optical counterpart of the radio transmission model outlined in Section 2 will have to decide on the relevance of a variety of published, groundbased, optical studies. This could be complicated by the need to evaluate whether the astronomers involved in a given study would have noticed an unexpected emission line appearing at an unexpected wavelength. Because many relevant astronomers are still alive, the suggested paper could include a unique feature—personal interviews.

## 7. Spaceships to Earth?

There is an additional method to investigate the possible presence of an ETI in the solar vicinity. Here we consider an ETI that went past our solar system at some time during the past 2 billion years, not one that is nearby now. This passage took a few million years. Suppose that, with its large space telescopes, a communicative ETI discovers that Earth is a living world with oceans and continents. What happens next? The ETI might attempt EM communication, say for millennia, but, if no reply is forthcoming, then ultimately it will face a choice: either abandon the effort (for a million years) or send a spacecraft.

Everything we know about human nature and history indicates that technological creatures will follow the latter path (e.g., B. Zuckerman 2002). Exploration of our solar system began with telescopic observations from Earth. But as soon as we developed the capability, we launched spaceships to explore planets and moons up close and personal.

Freeman Dyson expressed these thoughts about interstellar travel: "The problem of interstellar travel is a problem of motivation and not of physics" (in "Essays in Honor of Hans A. Bethe"; F. J. Dyson 1966), and "Interstellar distances are no barrier to a species which has millions of years at its disposal" (in "Disturbing the Universe"; F. J. Dyson 1979).

Consider a sustainable civilization that lasts for a billion years and passes 100 lt-yr from Earth. For travel over 100 lt-yr at velocity 1% of the speed of light (c), a trip by spaceship will take 10,000 yr. This trip time is $10^{-5}$ times the lifetime of the civilization. By analogy, if a person lives 100 yr, then the trip would take one-third of a day.

Various propulsion systems have been suggested to reach a velocity of $0.01c$. Some rely on onboard thermonuclear energy (R. Freeland 2013), and some on external energy sources, e.g., lasers or pellet streams (I. Crawford 1995; C. Singer 1995). The latter enable acceleration to high velocity without a need for onboard fuel. Freeland proposes a hybrid fission–fusion rocket where lithium deuteride (Li-D) pellets play the crucial role; the lithium absorbs neutrons and thus breeds the tritium needed for D–T fusion reactions.

Deceleration at the destination will always be a major problem, although a slow pellet stream could aid in braking (C. Singer 1995). A hybrid rocket accelerated by, say, a pellet stream and decelerated mainly by nuclear power might be optimal. Given how little we know about the construction of rocket propulsion systems—other than chemical and, perhaps, thermoelectric—it is plausible that a venerable technological civilization will have figured out how to travel at $0.01c$. If not, then the 100 lt-yr distance we assume for travel by spaceship might need to be reduced accordingly.

Earth, thanks to life, has had an oxygenic atmosphere for about 2 billion years (e.g., G. J. Cooke et al. 2025). Any ETI that possessed large space telescopes and that passed near our Sun during those years would have discovered that Earth is a living world. The number of Sun-like stars that passed within $R$ (= 100 lt-yr) of our solar system during the past 2 billion years is given by

$$nV\pi R^2 T,$$

where $V = 48$ km s$^{-1}$ (B. Anguiano et al. 2020; B. M. S. Hansen & B. Zuckerman 2021) is the typical relative velocity of stars in the solar vicinity, and $n = 3.6 \times 10^{-4}$ is the number of old solar-type stars per cubic light year (J. D. Kirkpatrick et al. 2024). Multiplying by T, the number of seconds in 2 billion years, one finds that about 2 million old, single, solar-type stars have passed within 100 lt-yr of our solar system. If any of these stars hosted an ETI, then, having discovered with their large space interferometers that Earth is a living world, they could have sent either a probe or themselves to our solar system. The most logical place for a probe would be in Earth-orbit high enough above our atmosphere to remain in orbit for a cosmically long time.

The absence of any Earth-orbiting probe—or the ETI themselves—is an indicator that none of the 2 million oldish, single, solar-like, stars that have passed within 100 lt-yr of Earth since its atmosphere became oxygenated harbored a curious, communicative ETI.

## 8. The Number of Communicative Technological Civilizations in the Milky Way ($N$)

A topic of great interest is "$N$," the number of technological civilizations existing in our Milky Way Galaxy that are interested in interstellar communication. Based on the model outlined in Section 2, it is (or will be) possible to place limits on $N$. For EM communication, we considered a volume of space of radius 200 pc and argued that an ETI within this volume, and seriously interested in interstellar communication, would be motivated to transmit continuously toward some hundreds of other solar-like (F6–M2) stars, including Earth. If so, then, after we have observed all the old solar-like stars in this volume without a positive result—this may already have been done via the totality of SETI and non-SETI radio plus optical wavelength surveys—we can conclude that no such star is orbited by a planet that hosts an ETI.

According to the 20 pc survey (J. D. Kirkpatrick et al. 2024), about 30% of the stars in the solar neighborhood are of spectral type F6–M2. If we assume 200 billion stars total in the Milky Way, and 100 billion are old, then 30,000,000,000 are old and of solar type. Of these, about 20,000,000,000 are single. This is to be compared to ~200,000 within 200 pc. Thus, a fraction $10^{-5}$ of old, solar-type star systems do not contain an ETI. If $N \sim 10^5$, then we might have expected one ETI within 200 pc to be transmitting toward Earth—this suggests an upper limit to $N$ of order 100,000.

As noted in Section 7, during the past 2 billion years, 2,000,000 single, solar-type stars sufficiently old (>4.5 billion years) to host an ETI have passed within 100 lt-yr of Earth. As in the preceding paragraph, the 2,000,000 is to be compared with the 20,000,000,000 old, single, solar-type stars in the entire Milky Way. If we assume that a purposely communicative ETI hosted by any of these 2 million Earth-passing stars would have sent a probe or themselves to our solar system, then a fraction $10^{-4}$ of the old solar-type star systems do not contain an ETI. If $N \sim 10^4$, then we might have expected one ETI to have sent a spaceship to Earth—this suggests an upper limit to $N$ of order 10,000.

While the above might seem like not very tight limits on $N$, they are smaller, substantially so, than many guesstimates for $N$ that have appeared in the literature during the past 50+ yr.

Furthermore, these upper limits to $N$ are based on the observational efforts of a wide variety of researchers dating back many decades, rather than on simple guesswork.

## 9. What about AI?

AI introduces a wild card into considerations of what a technologically advanced species might be motivated to do. Three possibilities are as follows:

(1) AI is the helper of a biological species that retains control.
(2) Equal but parallel paths of evolution where the interests of both the biological and AI entities continue to play a substantial role.
(3) AI takes over a world and Darwinian biology is insignificant.

If the biological creator of AI can train the AI to be "conscious"—say, similar to the computer "Hal" in the 1968 movie "2001: A Space Odyssey,"—and to experience emotions, including curiosity, then even for the third path, interest in interstellar communication might well persist.

## 10. Conclusions

Because radio SETI programs state their sensitivities in terms of isotropic radiated power, the implicit implication is that Earth has not been targeted in preference to other stars or directions. Thus, received signals may be faint. Furthermore, most of the stars covered in such programs are located well beyond the 200 pc distance we have considered here. ETIs at such distant stars will, typically, not be on the lookout for a "reply" from Earth and round-trip communication times, if somehow communication can be established, will be very long. In total, radio SETI surveys have observed only a modest percentage of the few hundred thousand relevant nearby stars that should be observed.

In contrast, numerous non-SETI surveys at radio and optical wavelengths (Sections 5 and 6) could have detected the continuous, strong signals that could be sent toward Earth by a nearby, purposely communicative, technological civilization. Broadband, general, radio sky surveys have covered essentially the entire sky, albeit over limited ranges of wavelengths. Similarly, the ensemble of conventional optical astronomical studies dating back 100 yr have covered the entire sky, probably with sufficient sensitivity to have detected the model communicative ETI described in Section 2. A paper that analyzes this history would be very worthwhile and, as noted in Section 6, could include some unique elements.

Notwithstanding the huge signals that a nearby ETI could beam continuously toward Earth, as motivated by the model described in Section 2, because the ETI could be transmitting at radio, infrared, or optical wavelengths, it would be premature to claim that we can now place a quantitative limit on the existence of such ETIs. The analysis suggested in the preceding paragraph, and also in Section 6, will probably show that no ETIs that transmit at optical wavelengths exist within a few hundred parsecs of Earth. But most of the radio band is still unexplored, and the same holds, even more so, for infrared wavelengths.

Complete *broadband* surveys at radio, infrared, and optical wavelengths of old solar-like stars in the solar vicinity could provide a definitive answer to the question of whether any purposely communicative ETI exists at any of these stars. In addition, the absence of any alien spacecraft probes in orbit around Earth, or the aliens themselves anywhere in the solar system, provides strong evidence that no ETI has passed closely by Earth during the past 2 billion years. A combination of such techniques can yield the first quantitative estimate (upper limit <100,000, or even <10,000) for "*N*," the number of communicative ETIs in our Milky Way Galaxy.

We are especially grateful to G. Marcy for numerous illuminating conversations. R. Freeland, I. Crawford, G. Matloff, D. Bardalez Gagliuffi, U. Bernstein, R. Modlin, and R. Schneider provided helpful advice. M. Morris, J. Larkin, and B. Hansen kindly read and furnished comments on a draft of this paper.

## ORCID iDs

B. Zuckermana https://orcid.org/0000-0001-6809-3045

## References

Abdurro'uf, Accetta, K., Aerts, C., et al. 2022, ApJS, 259, 35 Anguiano, B., Majewski, S., Hayes, C., et al. 2020, AJ, 160, 43
Becker, R. H., White, R. I., & Helfand, D. J. 1995, ApJ, 450, 559
Betz, A. 1986, AcAau, 13, 623
Bryson, S., Kunimoto, M., Kopparapu, R., et al. 2021, AJ, 161, 36
Cocconi, G., & Morrrison, P. 1959, Natur, 184, 844 Condon, J. J., Cotton, W. D., Greisen, E. W., et al. 1998, AJ, 115, 1693 Cooke, G. J., Marsh, D. R., Walsh, C., Sainsbury-Martinez, F., & Braam., M. 2026, CliPa, 22, 483
Crawford, I. 1995, in Extraterrestrials: Where Are They?, ed. B. Zuckerman & M. H. Hart (Cambridge Univ. Press),50
Dyson, F. J. 1966, in Perspectives in Modern Physics: Essays in Honor of Hans A. Bethe on the Occasion of His 60th Birthday, ed. R. E. Marshak (Interscience Publishers)
Dyson, F. J. 1979, Disturbing the Universe (Basic Books)
Fields, B., & Goodman, J. C. 2025, ApJ, in press (arXiv:2508.08628) Freeland, R. 2013, JBIS, 66, 290
Gillon, M., Ducrot, E., Bell, T. J., et al. 2026, NatAs, in press (arXiv:2509.02128)
Gray, R. H. 2021, JAHH, 24, 981
Guimond, C. M., Spohn, T., Berdyugina, S., et al. 2026, SSRv, 222, 8
Hansen, B. M. S., & Zuckerman, B. 2021, AJ, 161, 145

Hasegawa, Y., Dressing, C., & Carone, L. 2025, in HWO25
Howard, A. W., Horowitz, P., Wilkinson, D., et al. 2004, ApJ, 613, 1270
Isaacson, H., Siemion, A. P. V., Marcy, G. W., et al. 2017, PASP, 129, 054501
Kardashev, N. S. 1964, SvA, 8, 217
Kirkpatrick, J. D., Marocco, F., Gelino, C., et al. 2024, ApJS, 271, 55
Kollmeier, J., Rix, H.-W., Aerts, C., et al. 2026, AJ, 171, 52
Lacy, M., Baum, S. A., Chandler, C. J., et al. 2020, PASP, 132, 035001 Laskar, J., Joutel, F., & Robutel, P. 1993, Natur, 361, 615
Li, X., Wong, S., Ma, J., et al. 2025, ApJS, 281, 13
Lustig-Yaeger, J., Meadows, V. S., Tovar Mendosa, G., et al. 2018, AJ, 156, 301
Manunza, L., Vendrame, A., Pizzuto, L., et al. 2025, AcAau, 233, 155
Marcy, G. W., & Tellis, N. K. 2023, MNRAS, 520, 2121
Marcy, G. W., & Tellis, N. K. 2024, MNRAS, 531, 2669
Marcy, G. W., Tellis, N. K., & Wishnow, E. H. 2022, MNRAS, 515, 3898
Margot, J. L., Li, M. G., Pinchuk, P., et al. 2023, AJ, 166, 206
Matthews, L. D. 2025, PASP, 137, 116001
Miozzi, F., Shahar, A., Young, E. D., et al. 2025, Natur, 648, 551
Murphy, T., Mauch, T., Green, A., et al. 2007, MNRAS, 382, 382
Pecaut, M. J., & Mamajek, E. E. 2013, ApJS, 208, 9
Pezzotti, C., Betrisey, J., Buldgen, G., et al. 2026, A&A, 706, 257
Price, D. C., Enriquez, J. E., Brzycki, B., et al. 2020, AJ, 159, 86
Quanz, S. P., Ottiger, M., Fontanet, E., et al. 2022, A&A, 664, A21
Rogers, J. G., Schlichting, H. E., & Young, E. D. 2024, ApJ, 970, 47
Rugheimer, S., Segura, A., Kaltenegger, L., & Sasselov, D. 2015, ApJ, 806, 137
Ryan, D. J., & Robinson, T. D. 2022, PSJ, 3, 33
Schwartz, R. N., & Townes, C. H. 1961, Natur, 190, 205
Segura, A., Walkowicz, L. M., Meadows, V., Kasting, J., & Hawley, S. 2010, AsBio, 10, 751
Singer, C. 1995, in Extraterrestrials: Where Are They?, ed. B. Zuckerman & M. H. Hart (Cambridge Univ. Press),70 Soliz, J. J., & Welsh, W. F. 2026, arXiv:2601.02548 Tarter, J. 2001, ARAA, 39, 511
Townes, C. H. 1983, PNAS, 80, 1147
Tusay, N., Sheikh, S. Z., Sneed, E., et al. 2024, AJ, 168, 283 van Belle, G. T., Shaklan, S. B., & Kulkarni, S. R. 2025, Astronomical Optical Interferometry from the Lunar Surface, W. M. Keck Institute for Space Studies (KISS), Calif. Inst. of Tech.
Van Eylen, V., Massey, R., Awan, S., et al. 2025, arXiv:2512.16416
Wlodarczyk-Sroka, B. S., Garrett, M. A., & Siemion, A. P. V. 2020, MNRAS, 498, 5720
York, D. G., Adelman, J., Anderson, J. E., et al. 2000, AJ, 120, 1579
Young, E. D., Shahar, A., & Schlichting, H. E. 2023, Natur, 616, 306
Zink, J. K., & Hansen, B. M. S. 2019, MNRAS, 487, 246
Zuckerman, B. 1985, AcAau, 12, 127
Zuckerman, B. 2002, Mercury, 31, 14 (also arXiv:1912.08386)